\documentclass[useAMS,usenatbib,usegraphicx]{mn2e}

\title[Intense Metal Accretion at White Dwarfs]{Scars of Intense Accretion Episodes at Metal-Rich White Dwarfs}

\author[J. Farihi et al.]{J. Farihi$^1$, B. T. G\"ansicke$^2$, M. C. Wyatt$^3$, J. Girven$^2$, J. E. Pringle$^{1,3}$, A. R. King$^1$\\
$^1$Department of Physics \& Astronomy, University of Leicester, Leicester LE1 7RH, UK; jf123@star.le.ac.uk\\
$^2$Department of Physics, University of Warwick, Coventry CV4 7AL, UK\\
$^3$Institute of Astronomy, University of Cambridge, Cambridge CB3 0HA}

\begin{document}

\date{}

\maketitle

\label{firstpage}

\begin{abstract}

A re-evaluation of time-averaged accretion rates at DBZ-type white dwarfs points to historical, time-averaged rates 
significantly higher than the currently observed episodes at their DAZ counterparts.  The difference between the ongoing,
instantaneous accretion rates witnessed at DAZ white dwarfs, which often exceed $10^8$\,g\,s$^{-1}$, and those inferred 
over the past $10^5-10^6$\,yr for the DBZ stars can be a few orders of magnitude, and therefore must result from high-rate 
episodes of tens to hundreds of years so they remain undetected to date.  This paper explores the likelihood that such brief, 
intense accretion episodes of gas-phase material can account for existing data.  For reasonable assumptions about the 
circumstellar gas, accretion rates approaching or exceeding $10^{15}$\,g\,s$^{-1}$ are possible, similar to rates observed 
in quiescent cataclysmic variables, and potentially detectable with future x-ray missions or wide-field monitoring facilities.  
Gaseous debris that is prone to such rapid accretion may be abundant immediately following a tidal disruption event via 
collisions and sublimation, or if additional bodies impinge upon an extant disk.  Particulate disk matter accretes at or near 
the Poynting-Robertson drag rate for long periods between gas-producing events, consistent with rates inferred for dusty 
DAZ white dwarfs.  In this picture, warm DAZ stars without infrared excesses have rates consistent with accretion from 
particulate disks that remain undetected.  This overall picture has implications for quasi-steady state models of accretion 
and the derived chemical composition of asteroidal debris in DBZ white dwarfs.

\end{abstract}


\section{INTRODUCTION}

Atmospheric metals should not be present in isolated white dwarf stars with effective temperatures below
roughly 25\,000\,K.  At this stage of their evolution, radiative forces become insignificant \citep{cha95}, and 
gravitational settling pulls elements heavier than helium into the stellar interior on timescales a few to several 
orders of magnitude shorter than their cooling ages \citep{fon79}.  Nevertheless, white dwarfs in this temperature
range exhibiting atmospheric metal absorption have been known for nearly a century, and the prototype is still 
the nearest single white dwarf known, vMa\,2 \citep{van17}.  At 4.4\,pc, this star exemplifies a few characteristics 
of the metal pollution phenomenon: it is a single star and thus not accreting from a stellar or substellar companion 
\citep{far08a}, its location within the Local Bubble and helium-dominated atmosphere \citep{duf07} precludes 
accretion from interstellar material, and its age of 3\,Gyr is at odds with the persistence of metals that sink within
a few Myr \citep{koe09}.  Combined with the above facts, the abundance patterns in stars like vMa\,2 have long
suggested recent accretion from material such as that found in asteroids \citep{sio90,gra90}.

Indeed, evidence is now strong that metal-polluted white dwarfs accrete from planetary debris.  {\em Spitzer} 
studies have revealed circumstellar disks at a significant fraction of these stars, where the temperature profile and 
inferred geometry of the dust is consistent with material completely contained within the Roche limit for km-sized 
or larger solid bodies, with innermost disk temperatures that should rapidly sublimate solids, and drive material 
onto the star \citep{xu11,far09,jur07,von07,gan06,rea05}.  The favored model for the origin of this material is 
the tidal disruption of a large asteroid analog \citep{jur03}, perturbed into a high eccentricity by an unseen body 
such as a major planet \citep{deb12,bon11,deb02}.  All available mid-infrared spectroscopy reveals disk matter 
that is silicate-rich and carbon-poor \citep{jur09,rea09}, while stellar spectroscopy reveals atmospheric pollution 
by refractory-rich and volatile-poor material \citep{far10a,jur06}.  Critically, high-resolution optical spectroscopy of 
the disk-contaminated stars reveals distinctly terrestrial-like abundances \citep{kle10,zuc07}.  Thus metal-enriched 
white dwarfs are astrophysical traps for exoplanetary debris, acting as detectors that can yield the bulk composition 
of planetary building blocks that orbit intermediate mass stars.

This paper re-examines in detail the inferred accretion rates at both hydrogen and helium atmosphere, cool white 
dwarfs with trace metals, referred to here as DAZ and DBZ stars respectively, based on standard spectral types.  As
described in \citet{gir12}, an improved calculation method is used to place both classes on equal footing, and in doing 
so, the derived accretion rates, averaged over the recent history for the DBZ stars, are found to be significantly higher 
than the current rates observed at the DAZ stars.  This work explores these issues in detail and develops a model to 
address the observations.  In \S2 the accretion rate calculations are described and examined, and \S3 presents a 
scenario that can account for these data.  Implications for future observations and the nature of the destroyed and 
accreted planetary bodies are discussed in \S3 and \S4.

\section{Inferred Accretion Rates for DAZ and DBZ Stars}

\subsection{Steady State Metal Accretion}

If a star is in a steady state between accretion and diffusion, the total mass accretion rate for heavy elements can 
be expressed as \citep{dup93}

\smallskip
\begin{equation}
\dot M_z = \sum\limits_{i} \frac{X_{i}M_{\rm cvz}}{\tau_i}
\label{eqn1}
\end{equation}

\smallskip
\noindent
where $X_i$ is the mass fraction, $\tau_i$ the diffusion (i.e., sinking) timescale for the $i^{\rm th}$ heavy element, 
and $M_{\rm cvz}$ is the mass of the stellar convection zone or outer atmosphere (defined at the base of the convection 
zone or at Rosseland optical depth 5, whichever is deeper; \citealt{koe09}).  The numerator on the right-hand represents 
the mass of a the $i^{\rm th}$ element in the convective or outer layers of the star.

All known metal-polluted white dwarfs exhibit the Ca\,{\sc ii} K line in optical spectroscopy owing to the strength 
of this atomic transition at this range of effective temperatures (for this reason it is the strongest absorption feature 
in the Sun).  Accretion rate calculations have been naturally tied to this element, and one can derive an accurate 
infall rate for calcium, and extrapolate to the total mass accretion rate with an appropriate correction factor, as 
follows  

\smallskip
\begin{equation}
\dot M_z \approx \frac{1}{A} \frac{X_{\rm Ca}M_{\rm cvz}}{\tau_{\rm Ca}}
\label{eqn2}
\end{equation}

Prior calculations assumed the infalling material was exactly solar (i.e., including hydrogen and helium; 
\citealt{koe06}), or 1\% solar by mass (i.e., no hydrogen or helium; \citealt{jur07}); the latter yields $A=1/109.1$
or $1/42.7$ depending on whether the extrapolation is made from a calcium abundance measured as [Ca/H] or 
[Ca/He] respectively \citep{lod03}.

\subsection{Instantaneous vs. Historical Rates}

For $T_{\rm eff}\ga11\,000$\,K DAZ white dwarfs, Equation \ref{eqn1} can be applied with confidence because 
these stars possess tiny convection zones or none at all, and have commensurately short metal sinking timescales 
(on the order of days; \citealt{koe09}) that imply a steady state is virtually certain.  Rates computed in this way for 
warm DAZ stars are ongoing and {\em instantaneous} measures of the infall of a given heavy element.  In contrast, 
for the coolest DAZ stars and the DBZ stars in general, the assumption of a steady state is uncertain owing to larger 
diffusion timescales from correspondingly deeper convection zones \citep{paq86}.  The DBZ stars can retain 
atmospheric metals for timescales up to $10^5-10^6$\,yr, and while they cannot yield instantaneous accretion 
rates, their sizable convection zones contain the integrated signatures of accretion; a substantial mass of metals 
acquired within the last few sinking timescales.  Because of this, DBZ stars provide a historical record of accretion 
and, importantly, a minimum mass for the destroyed parent body or bodies \citep{kle10,far10a,jur06}.

In order to best compare these two classes of polluted stars, \citet{far09} calculated {\em historical} accretion rates 
for DBZ stars; that is, time-averaged over a single diffusion timescale, assuming a steady state.  The likelihood of 
an infrared excess due to circumstellar dust at DAZ stars is correlated with ongoing, high-rate accretion, and thus 
a similar correlation in the DBZ stars was naturally sought.  While the fraction of stars with infrared excess differs 
between the DAZ and DBZ white dwarfs with accretion rates above a benchmark of $3\times10^8$\,g\,s$^{-1}$ 
\citep{gir12,xu11,far10b,far09}, the relevant point for this study is that until recently, in terms of metal pollution alone 
and based on the assumption of solar calcium relative to other heavy elements, the two stellar classes appeared to 
have broadly similar, inferred metal accretion rates.  

In what follows, the terms {\em instantaneous} and {\em ongoing} are used to refer to accretion rates for warm DAZ 
stars where a steady state is highly probable, while {\em historical} and {\em time-averaged} refer to the rates derived 
for DBZ stars and the coolest DAZ stars, where an integrated approach is necessary.

\begin{figure}
\includegraphics[width=86mm]{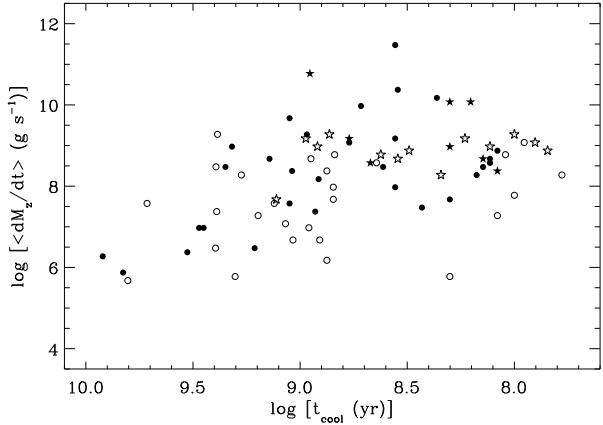}
\caption{Total metal accretion rates onto 78 contaminated stars observed with {\em Spitzer} IRAC.  The open 
and filled symbols respectively denote DAZ and DBZ-type white dwarfs, while the stars and circles respectively 
distinguish those white dwarfs with and without infrared excesses due to circumstellar dust.  Among these targets 
alone there are 8 DBZ stars with accretion rates above the maximum witnessed to be ongoing for any known DAZ 
white dwarf.  Cooling ages are calculated based on white dwarf atmospheric and evolutionary models \citep{fon01}.
\label{fig1}}
\end{figure}

\subsection{New Accretion Rate Calculations}

Assuming 0.01 solar composition by mass, extrapolated from calcium abundance, for both DAZ and DBZ white 
dwarfs is not ideal, but understandable based on existing data just a couple years prior.  However, as discussed 
in \citet{gir12}, a more accurate estimation of accretion rates is possible based on recent work.  Briefly, there are 
now eight  stars with measured O, Mg, Si, Ca, and Fe\footnote{These elements combined represent over 95\% of 
the bulk Earth \citep{all95}} abundances, with some of these stars exhibiting further elements such as Al, Ti, Cr, and 
Ni \citep{gan12,kle11,mel11,far11a,zuc10,ven10,duf10,kle10}.  These data demonstrate that the accreted matter in 
polluted white dwarfs is broadly similar to rocky material of the inner Solar System.  Specifically, \citet{zuc10} find 
that calcium represents, on average, close to 1:60 by mass, of all the accreted heavy elements for well-studied 
stars, while this same ratio is 1:62.5 for the bulk Earth \citep{all95}.

\begin{figure}
\includegraphics[width=86mm]{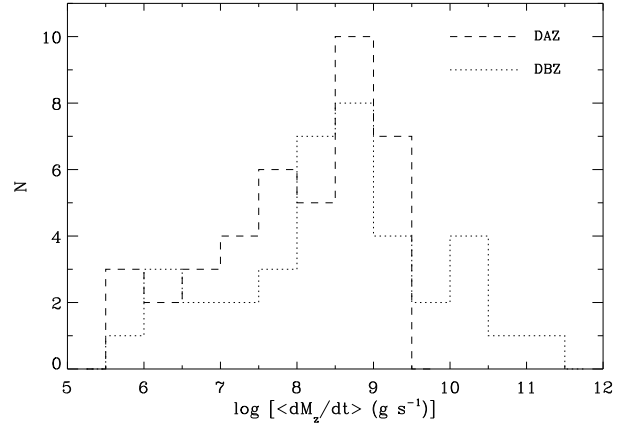}
\caption{Histogram of the rates plotted in Figure \ref{fig1} for each atmospheric class of polluted white dwarf.  The
population of rates $\log \langle dM_z/dt\,({\rm g\,s})^{-1} \rangle > 9.5$ occur only among the DBZ-type stars.
\label{fig2}}
\end{figure}

These results suggest an improved metal accretion rate for dozens of stars where only calcium is detected can 
be made assuming it is 1.6\% ($A=1/62.5$) of the total mass.  Figure \ref{fig1} plots accretion rates calculated 
using Equation \ref{eqn2} for 40 DAZ and 38 DBZ stars observed with {\em Spitzer} IRAC \citep{gir12,far10b,far09,
far08b,jur07}.  Calcium abundances and stellar parameters come from the most recent studies (e.g., \citealt{koe06}),
with sinking timescales and convective envelope masses given in \citet{koe09}.  Comparing this plot with those 
published previously, it is clear there are an increasing number of outlying DBZ-type stars with inferred rates one 
to a few orders of magnitude above that observed at any DAZ star.  Figure \ref{fig2} re-plots these accretion rates 
as histograms for both atmospheric types (ignoring dust emission), clearly delineating a subgroup of high values that 
occur exclusively for DBZ stars.  While GALEX\,1931$+$0117 is currently the most extreme example of a DAZ star, 
with an instantaneous accretion rate of $2\times10^9$\,g\,s$^{-1}$ \citep{gan12}, its extreme DBZ-type counterpart is 
HE\,0446$-$2531 which has a time-averaged rate two orders of magnitude higher at $3\times10^{11}$\,g\,s$^{-1}$.  
Figures \ref{fig1} and \ref{fig2} plot a large subset of all nearby (pre-SDSS) metal-enriched white dwarfs, but there 
are additional examples of high rate DBZ stars (e.g., Figure 11 of \citealt{xu11}), and at least 10 cases where the 
historical accretion rates are higher than any inferred to be ongoing at DAZ stars.  

Table \ref{tbl1} compares accretion rate calculations for the eight stars with robust detections and abundances of 
O, Mg, Si, Ca, and Fe: in the second column using Equation \ref{eqn2} and assuming calcium is 1.6\% of the total 
accreted metal mass (as described above), and in the third column from the actual detected elements using Equation 
\ref{eqn1}.  While a few of these stars exhibit additional elements beyond the five in common, the third column values 
represent a minimum, as the detection of further elements (e.g., Al, Ni) may serve to increase the calculated rates.  
However, because most of the major terrestrial elements are detected in these stars, it is possible but unlikely their 
total accretion rates will change substantially.  Comparing the inferred rates from both methods, there is agreement 
to within $\pm0.2$\,dex for five of the eight stars and overall, while three are over-predicted.  These over-predictions 
may diminish as further elements are detected in these stars, but it is more likely that the (modest) disagreement in 
the calculated accretion rates are due to real metal-to-metal abundance variations among the parent bodies now 
falling onto the stars as debris \citep{zuc11,kle10}.

\subsection{Potential Biases}

It is worth noting that GALEX\,1931 is the only known DAZ star with {\em optical} detections of all the necessary 
metals for the Table \ref{tbl1} comparison.  This is because, all else being equal, the atmospheric opacity of hydrogen 
is significantly higher than that of helium \citep{zuc03}.  Because of this, there is a bias against the optical detection 
of DAZ stars with modest to low amounts of pollution (see Figure 1 of \citealt{koe06} for an excellent illustration of this 
fact).  Within the DA cooling sequence, the combination of atmospheric opacity and Ca\,{\sc ii} ionization fraction permits 
lower metal abundances and accretion rates to be detected at cooler stars.  But between the two classes, the detection 
of metal lines in DA stars is more difficult, favoring those metal-rich stars exhibiting relatively strong lines, and thus high 
abundances and accretion rates \citep{koe05}.  {\em Because the bias works in the opposite direction, the lack of high 
rates in the DAZ population is real}.

Another possibility is that more vigorous accretion takes place at DBZ stars.  However, it is difficult to conjure any
scenario that physically favors higher rate accretion of planetary debris at stars with helium atmospheres.  There 
is neither empirical nor theoretical reasoning that suggests the existence of DB stars is related to anything beyond 
(single) stellar evolution: a very late thermal pulse, immediately following the asymptotic giant phase, can enrich 
the outer layers of a star with hydrogen-deficient material \citep{her99}, eventually producing a helium-dominated 
remnant \citep{wer94}.  Other explanations along these lines would include that DB stars tend to have planetary 
system architectures more efficient at asteroid perturbation, or parent bodies highly enriched in calcium, but such
solutions are contrived and therefore discounted.

Lastly, white dwarf atmospheric models do not have substantial uncertainty in the sizes of the convective or 
outer envelopes, nor in the gravitational settling behavior at the bottom of these zones \citep{koe09}.  However, 
it is feasible that there are gaps in the theoretical understanding of stellar atmospheres that may contribute to the 
inferred trends, e.g., if the settling times in Equation \ref{eqn1} are systematically under-predicted for DB stars.  
This possibility is outside of the scope of this paper and not discussed further.

The following sections proceed on the assumption that a planetary system forms without any foreknowledge 
of the ultimate stellar remnant, and that the planetary bodies within depend only on the conditions at formation 
and the subsequent influence of stellar evolution and mass loss on those bodies (i.e., only stellar radiation and 
dynamical changes).

\begin{table}
\begin{center}
\caption{Comparison of Accretion Rate Estimates For Stars with Detected O, Mg, Si, Ca, Fe\label{tbl1}} 
\begin{tabular}{@{}lccc@{}}
\hline

Star				&$\log \langle dM_z/dt ({\rm g\,s}^{-1}) \rangle^a$		
				&$\log \langle dM_z/dt ({\rm g\,s}^{-1}) \rangle^b$\\
\hline

G241-6			&9.8			&9.3\\
GALEX\,1931		&9.2			&9.2\\
GD\,40			&10.1		&9.4\\
GD\,61			&9.0			&8.8\\
HS\,2253$+$8023	&10.2		&10.1\\
PG\,1015$+$161	&8.5			&8.2\\
SDSS\,0738		&10.3		&10.3\\
SDSS\,1228		&9.0			&8.8\\

\hline

Log Average		&8.66			&8.43\\

\hline

\end{tabular}
\end{center}

$^a$ Assuming Ca is 0.016 of the total accreted mass.\\
$^b$ From measured abundances of at least O, Mg, Si, Ca, Fe \citep{gan12,kle11,far11a,mel10,ven10,kle10,duf10,zuc10}.\\

{\em Note}.  The overall differences in the inferred accretion rates between the two columns are minor, and will 
not affect the findings for DBZ versus DAZ properties; see \S 2.3 for calculation details.  

\end{table}

\section{A History of High Accretion Rates}

The inescapable conclusion of the data shown in Figure \ref{fig2} is that there must be relatively short-lived 
episodes of accretion at metal-polluted white dwarfs; brief enough not to have been observed as ongoing, and 
at rates substantially higher than those witnessed to date at DAZ stars.  This is obviously true for the six stars 
with time-averaged rates above $10^{10}$\,g\,s$^{-1}$, but it also appears likely for the greater DBZ population. 
Comparing the rates between the DAZ and DBZ groups plotted in the figures, the highest accretion rates differ 
by a factor of 160, the average of the top six rates differ by a factor of 40, and the average of all the rates differ 
by a factor of 25.  Thus the phenomenon is represented by the entire class of helium atmosphere stars, and 
accentuated by a handful of outstanding examples.

The rates inferred for the DBZ stars are the mass of accreted metals over a single diffusion timescale, divided by 
that timescale; around $10^5$\,yr at 17\,000\,K and $10^6$\,yr near 12\,000\,K \citep{koe09}.  However, in cases
where circumstellar dust is not detected at a given DBZ white dwarf, it is possible that accretion halted within the
past one to several diffusion timescales.  In this sense, the rates inferred for the DBZ stars are {\em minimum}, 
time-averaged rates, as atmospheric metal abundances will exponentially diminish, but remain potentially detectable 
for a few to several sinking timescales after accretion has ended.  For simplicity, the following assumes all observed
metals are accreted over a single diffusion timescale; if accretion is no longer ongoing, then prior accretion must have 
occurred at even higher rates than those inferred below.

\subsection{High and Low State Model}

For a toy model with two distinct and constant accretion rates -- a high state and a low state -- taking place over a 
single diffusion timescale for a typical DBZ star, the observed mass of metals in the stellar convection zone $M_z$ 
can be written as:

\smallskip
\begin{equation}
\dot M_z \tau = \dot M_ht_h + \dot M_lt_l
\label{eqn3}
\end{equation}

\smallskip
\noindent
where $t_h/\tau$ is the fraction of time the star is accreting at the high rate $\dot M_h$, similarly $t_l/\tau$ is the fraction 
of time spent accreting at the low rate $\dot M_l$, and where $\tau$ is the diffusion timescale.  The right hand side of 
Equation \ref{eqn3} simply breaks the total accreted mass into that delivered in two distinct states, but one can easily 
extend this to a continuum of rates and corresponding intervals; this does not change the conclusions reached here.  
As shown in \S3.2, since $t_h\ll t_l$, it follows that $t_l\sim\tau$ and thus $t_h/t_l$ is the fraction of the total time spent 
accreting in the high state.  This allows for the possibility that $\dot M_h$ occurs in a succession of episodes, not just 
one per DBZ diffusion timescale.  The mass represented by both sides of Equation \ref{eqn3} is typically $10^{22}$\,g 
for a given DBZ star, but this can vary by up to two orders of magnitude \citep{gir12,far10a}. 

Identifying $\tau$ and $\dot M_z$ as the diffusion timescale and time-averaged accretion rate for DBZ-type stars, 
and equating $\dot M_l$ to that measured for the DAZ white dwarfs, Equation \ref{eqn3} permits the assessment 
of ($t_h, \dot M_h$) values sufficient to account for the inferred differences among the instantaneous and historical 
accretion rates.  Figure \ref{fig3} plots three representative solutions for $\tau=10^6$\,yr, with parameters described 
in Table \ref{tbl2}.  From top to bottom are plotted the resulting ($t_h, \dot M_h$) for pairs of $\dot M_z$ and $\dot M_l$ 
representing 1) the single highest accretion rates in each class, 2) the average of the top six rates in each class, and 
3) the average among all stars in each class from Figure \ref{fig1}.  

\subsection{Allowed Timescales for High Rate Episodes}

Because ongoing accretion rates as high as those inferred for the DBZ have not been witnessed to date, strict limits 
can be placed on the timescales for such episodes, $t_h$.  For $\tau=10^6$\,yr and a high-rate accretion timescale 
of $t_h=10^5$\,yr ($0.11t_l$), the probability that at least one of 46 known DAZ stars will be accreting at the high rate 
is above 99\%, while for $t_h=10^4$\,yr ($0.01t_l$) this same probability drops to 37\%.  Based on this argument alone,
one can conclude $t_h\ll t_l$, and the timescale for any high-rate episodes must be no more than $10^4$\,yr.  Moreover, 
a few of the most highly polluted DBZ stars have $\tau\sim10^5$\,yr (e.g., GD\,362, Ton\,345), which supports even shorter 
timescales for high-rate spikes.

\begin{table}
\begin{center}
\caption{Accretion Rates for Figure 3\label{tbl2}} 
\begin{tabular}{@{}lccr@{}}
\hline

				&DBZ stars						&DAZ stars			&\\
Line				&$\log(\dot M_z[{\rm g\,s}^{-1}])$		
				&$\log(\dot M_l[{\rm g\,s}^{-1}])$							&Description\\

\hline

Dash			&11.5							&9.3					&Max $\dot M$\\
Long Dash		&10.8							&9.2					&Top 6 $\dot M$ \\
Solid				&10.1							&8.7					&Avg $\dot M$\\

\hline

\end{tabular}
\end{center}

{\em Note}.  There are six DBZ stars with inferred accretion rates above $10^{10}$\,g\,s$^{-1}$ (see Figure \ref{fig2}).

\end{table}

Notably, DA stars without detected metals constrain $t_h$ more strongly.  The high-resolution spectroscopic studies 
of \citet{zuc03} and \citet{koe05} collectively surveyed approximately 530 DA white dwarfs for photospheric metals, 
where stars more highly polluted than GALEX\,1931 would have been readily detected.  It is also likely that extreme
metal abundances in DA white dwarfs are identifiable in the low-resolution spectra of the Sloan Digital Sky Survey;
SDSS\,1228 is one of the most metal-rich DAZ stars and its discovery spectrum exhibits a clear Mg\,{\sc ii} absorption 
line at $g=16.2$\,AB\,mag \citep{gan06}.  While this star is relatively bright among the few thousand spectroscopically 
confirmed DA white dwarfs identified in SDSS DR7, there are nearly 470 DA stars with $g<17.0$\,AB\,mag where very 
high metal abundances (and thus accretion rates) were not detected  \citep{gir11}.  Based on these surveys, it is likely 
that $t_h<10^3$\,yr.

\subsection{Rapid Accretion of Gaseous Debris}

\begin{figure}
\includegraphics[width=86mm]{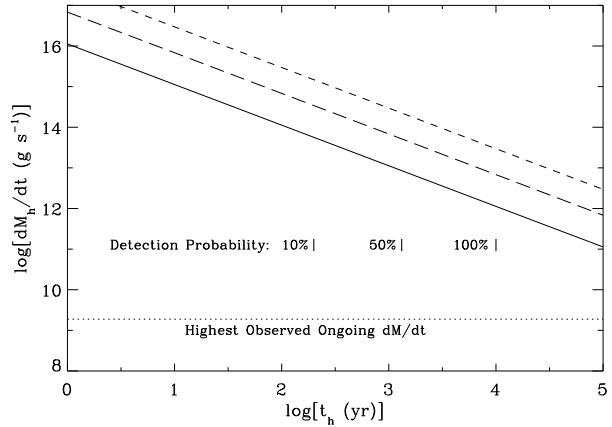}
\caption{Necessary ($t_h, \dot M_h$) that satisfy Equation \ref{eqn3} for $\tau=10^6$\,yr, and Table \ref{tbl2} values
for $\dot M_z$ and $\dot M_l$.  It should be noted that there is little difference in the plotted lines and those that result 
from $\dot M_l=0$.  Also shown are resulting benchmark probabilities of detecting high states of marked durations, 
given the sample of 534 DA stars surveyed at high resolution with Keck and VLT \citep{koe05,zuc03}.  Episodes of 
high rate accretion lasting tens to hundreds of years are allowed, while if the timescales approaches or exceeds 
$10^3$\,yr then the likelihood of detection at DAZ stars increases towards unity.  This result is corroborated by the
lack of extremely polluted DAZ stars in the SDSS DR7 \citep{gir11}.
\label{fig3}}
\end{figure}

Using the $\alpha$ prescription \citep{kin07} for completely gaseous accretion disks, \citet{jur08} estimates that a disk 
composed purely of metallic gas will dissipate within $10^5$\,yr, for $\alpha=0.001$.  While this timescale is longer than
the allowed range of $t_h$ in Figure \ref{fig3}, below it is shown that a reasonable choice of $\alpha$ can produce both 
$t_h$ and $\dot M_h$ values necessary to account for the accretion rate data.

The accretion rate of a completely gaseous disk can be expressed as

\smallskip
\begin{equation}
\dot M = \frac{m}{t_\nu}
\label{eqn4}
\end{equation}

\smallskip
\noindent
where $m$ is the mass of the disk and $t_\nu$ is the viscous timescale at a distance $d$ from the central star

\begin{equation}
t_\nu = \frac{d^2}{\nu}
\label{eqn5}
\end{equation}

\smallskip
\noindent
The viscosity $\nu$ can be rewritten following the $\alpha$ prescription \citep{pri81} and Equation \ref{eqn5} 
becomes

\begin{equation}
t_\nu = \frac{p}{\alpha} \left( \frac{d}{h} \right)^2
\label{eqn6}
\end{equation}

\smallskip
\noindent
where $p$ is the orbital period and $h$ is the scale height of the disk.  For an ideal, isothermal gas orbiting a star 
of mass $M$ at a distance $d$, the scale height is \citep{pri81}:

\begin{equation}
h = \sqrt{ \frac{kTd^3}{GM\mu} }
\label{eqn7}
\end{equation}

\noindent
where $T$ is the temperature of gas with mean molecular weight $\mu$.  

Metallic, circumstellar gas has been modeled at temperatures around 6000\,K in the few, well-studied stars with 
calcium triplet emission lines, using three independent means \citep{har11,mel10,gan06}. Gas composed of singly 
ionized O, Mg, Si, Fe will have a mean particle mass $\mu=2.5\times10^{-23}$\,g, and if this orbits at 0.5\,$R_{\odot}$ 
about a 0.6\,$M _{\odot}$ white dwarf, then $d/h\approx260$.  Then the viscous timescale at this distance becomes
$t_\nu\approx(10/\alpha$)\,yr and in the range 25 to 100\,yr for $\alpha=0.1-0.4$; typically required to account for 
observational data in a wide variety of fully ionized disks \citep{kin07}.  This viscous timescale is consistent with 
observations in two ways: 1) it is sufficiently short as to remain currently undetected at DAZ stars, and 2) for the 
largest, accreted metal masses in DBZ star convection zones ($10^{24}$\,g; \citealt{gir12,far10a}), it implies rates 
up to $10^{15}$\,g\,s$^{-1}$ (and higher for smaller $d$).

Metal-rich, pure gas disks at such temperatures will be highly or fully (singly) ionized as their overall ionization 
potential will be lower than similar accretion disks of solar abundance, hydrogen-dominated material\footnote{Of the 
expected major elements in a disk of vaporized rock, only O\,{\sc i} has an ionization potential comparable to H\,{\sc 
i}.}.  Thus, a disk of gaseous metals will be conductive, stable, and the standard $\alpha$ should hold \citep{sto00,
gam98}.  Accretion of gaseous debris via viscous dissipation with $\alpha=0.1-0.4$ thus nicely accounts for the data 
in Figure \ref{fig2}, yielding the correct range of rates and timescales dictated by Figure \ref{fig3}, and is hence plausible.  

Two recent studies have also shown that short bursts of high rate accretion are possible from disrupted asteroids at
white dwarfs.  \citet{raf11b} demonstrates that, under certain conditions, gas may efficiently couple with solids and lead 
to the inward drift of particulate disk matter at rates greatly exceeding that of Poynting-Robertson forces.  In this scenario, 
a spike of maximal accretion occurs at the very end of a disk episode \citep{met12}.  In contrast, \citet{bea12} show that a 
Nova-like, accretion-driven outburst may occur during the initial phases of an asteroid disruption and disk formation.  The 
standard $\alpha$ disk model used here applies to any epoch where sufficient gas is produced, and overlaps with these 
other models under appropriate conditions.

\subsection{Generating Asteroid-Sized Masses in Gas}

This section considers the efficient vaporization of significant masses of solid matter originally contained in extrasolar 
asteroids; i.e., processes able to reproduce the rates deduced in the previous sections.  Such gas production must be 
orders of magnitude more vigorous than sublimation of dust at the inner edge of a flat disk (see \S3.5), as this cannot 
produce the sufficiently high infall rates dictated by the accretion histories represented by Figure \ref{fig3}.  The following 
should be considered for heuristic purposes only.

One possibility is a planetesimal impinging on a pre-existing disk from the prior asteroid disruptions, as investigated 
in detail by \citet{jur08}.  In that model, if the incoming asteroid is less massive than the disk, its entire debris mass is 
reduced to gas via collisions and sputtering, while if it is more massive, then the disk mass is vaporized and the excess 
asteroid mass persists in solids.  Both cases have the potential to produce a large asteroid mass of gaseous debris and
allow for the possibility that a disk of solids will remain.  The metal-enriched white dwarfs G166-58 and PG\,1225$-$079 
both appear to have dust rings with enlarged inner holes \citep{far10b,far08b}, where an impact may have destroyed the 
inner region solids but left the outer disk intact.

A second scenario is a lone tidal disruption event.  A planetesimal will be shredded on its first approach, but the 
fragments will simply continue along their eccentric orbits (i.e., none will impact the star), dispersed but tightly centered 
around the original orbit of the parent body, analogous to stellar disruptions at black holes \citep{lod09,ros09}. The pieces 
will be sufficiently small as to be immune to tidal gravity, but will collide at periastron on subsequent passes.  Such collisions 
will occur at high relative velocity and so can generate gas \citep{tie94} as well as dust.  Furthermore, solid debris will be 
prone to sublimation while not yet shadowed by a disk, as optically thin material is heated above 1200\,K within $1.0\,R_
{\odot}$ (the approximate Roche limit) of a typical 15\,000\,K white dwarf.  These modes of gas production will eventually 
abate, as eccentric motion damping reduces collisional velocities and particles become shadowed by a growing (flat) 
disk of material.  The complete reduction of an asteroid into a disk will likely take many orbital periods; e.g., taking the 
11.8\,yr orbital period of Jupiter as a benchmark, one would expect planetesimals perturbed from orbits at similar 
distances from their host stars will take hundreds of years to make several passes.

The timescale for particle collisions in an optically thin disk within the Roche limit of a cool white dwarf is typically
shorter than Poynting-Robertson drag timescale by an order of magnitude (but with a linear dependence on optical 
depth and particle size; \citealt{far08a}).  In this regime, planetesimal debris generated by tidal disruption can only 
be reduced by collision or sublimation, at least until the disk becomes optically thick, at which point the disk evolution 
will be dominated by Poynting-Robertson drag at the inner edge, and viscous forces elsewhere \citep{raf11a}.  Disk 
observations that reveal 1) silicate emission features in the mid-infrared \citep{jur09,rea09} and 2) radially coincident 
gas and dust at some polluted white dwarfs \citep{far12,mel10,bri09,gan08,gan07,gan06} are consistent with a picture 
whereby the shattered parent body is at least partially (if not totally) reduced to micron-sized and smaller particles.  It is 
therefore plausible that a significant gaseous mass will result from the initial phases of a tidal disruption event.

\subsection{Observational Consequences and Predictions}

This rapid gas accretion scenario has implications for observations of circumstellar disks and metal-polluted white 
dwarfs.  Perhaps most notably, those stars suspected to be accreting gaseous debris {\em alone} should exhibit the 
highest rates, and far exceed the Poynting-Robertson drag rates for white dwarfs.  Ultimately, all material falling onto
the stellar surface will be in gas phase, and circumstellar disks may have 1) a cool outer region dominated by solids, 
2) a narrow transition region where solids and gas coexist, and 3) a hot inner region where all debris is sublimated.
Such architecture is inferred for the dust disks found and characterized by {\em Spitzer} (e.g., \citealt{far09,jur07,von07}, 
and the limiting factor in their infall rates is the (solid) mass per unit time delivered to the edge of the sublimation zone 
by Poynting-Robertson forces \citep{raf11a}.  Once vaporized, the model presented here predicts that gaseous accretion 
takes place on short timescales, but will produce very different infall rates for disks whose total mass is dominated by 
solids, and those dominated by gas.  

Based on the above, those disks detected to have spatially coincident dust and gas (six are known at present:
\citealt{bri12,duf12,mel12,far12,mel10,bri09}) are likely to be dominated by solids.  Critically, the best observations to 
date indicate that the (calcium) gas emission drops off sharply at the inner orbital radius where the particulate disk is 
modeled to terminate \citep{mel10,bri09}.  These data are consistent with the rapid depletion of a completely gaseous 
inner region, resulting in a relatively low surface density, a corresponding lack of detected emission, and are compatible 
with the model presented here.

\begin{figure}
\includegraphics[width=86mm]{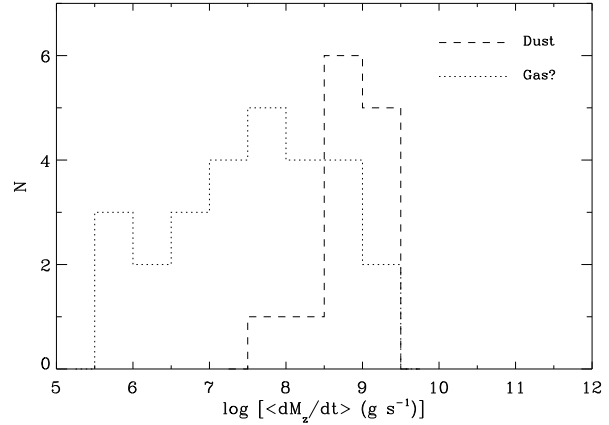}
\caption{Accretion rates for the 40 DAZ stars in Figure \ref{fig1}.  The dashed and dotted histograms show stars 
with and without an infrared excess, respectively.  Among 19 stars with $T_{\rm eff}\ga10\,000$K and where ongoing 
accretion is highly probable, 11 stars accreting from dust disks have an average infall rate more than three times higher 
than eight stars that lack infrared excesses.   If these latter stars accrete from gas-dominated circumstellar disks \citep{jur08}, 
the difference in inferred accretion rates would be contrary to the predictions made here, which suggest that pure gas disks 
achieve the highest accretion rates, whereas solid-dominated disks are limited by Poynting-Robertson forces \citep{raf11a}.
\label{fig4}}
\end{figure}

Notably, observations reveal that white dwarfs with detectable dust disks have, on average, the highest ongoing 
accretion rates.  Figure \ref{fig4} plots a histogram of metal accretion rates for DAZ stars with and without infrared 
excesses from circumstellar disks.  While essentially no stars exceed the range of infall rates expected for Poynting$
$-Robertson drag, the stars without obvious dust are the slowest accretors.  Many of the DAZ without infrared excess 
are sufficiently warm that they must be accreting unseen material; either gaseous or tenuous dust disks \citep{far10b}.  
The picture presented here predicts these stars are indeed accreting from undetected particulate disks, rather than 
purely gaseous matter, but it is also plausible that $\alpha$ varies substantially between gas disks \citep{jur08}.  
More sensitive infrared observations can test this prediction.

Detailed spectroscopy of DBZ white dwarfs has the potential to constrain the frequency of events that produce 
significant masses of gaseous debris, or distinct pollution events more generally.  For a given baseline of exo-chondritic 
or -terrestrial compositions, the heaviest elements such as iron and nickel should be under-abundant in DBZ stars if the 
timescale for major asteroid incursions is longer than the lifetime of photospheric metals.  To date there is a mix of 
observational results: there appear to be two white dwarfs polluted by intrinsically iron-poor material \citep{zuc11,far11a}, 
while most iron-deficient stars are consistent with gravitational settling \citep{koe11,far11b}.  Analysis of a large number 
of DBZ (and very cool DAZ) stars is needed to better understand these abundance patterns, as it remains possible that 
extrasolar, solid planetary material can be distinct from that contained in any known (or modeled) rocky bodies.  But it 
is plausible that certain abundance patterns can only be maintained if discrete, substantial pollution events occur on 
timescales shorter than the heavy element lifetimes in typical DBZ stars.

It is noteworthy that the toy model of high and low accretion states, which is certainly an approximation of reality, 
underscores the quasi-steady state accretion model recently outlined by \citep{jur12}.  In that model, the heavy element 
abundances observed in the atmospheres of DBZ-type white dwarfs may differ substantially from both the steady state
and early phase models \citep{koe09}.  Under certain conditions, parent body chemical abundances can be uncertain
by a factor of two in mass.  The results presented here constrain $r$, but not $\omega$ in Equation 10 of \citet{jur12}.

\subsection{Total Mass Accreted in High State}

An important consequence of the proposed scenario is the necessary mass (in gaseous heavy elements) delivered 
during the high state accretion episodes, and specifically the total mass accreted within $t_h$ so that its signature is 
still visible today among the DBZ population.  The required mass depends on the choice of constants in Equation 
\ref{eqn3}, but for $\tau=1$\,Myr a mass of $\dot M_h t_h \sim 10^{23}$\,g will reproduce the average DBZ accretion 
rates, for $\dot M_l$ equal to the average DAZ rate; this solution is represented by solid line in Figure \ref{fig3}.  If 
instead the diffusion timescale for the DBZ stars is taken to be $10^5$\,yr or smaller (as for GD\,362 and Ton\,345), 
then the necessary mass decreases to the order $10^{22}$\,g.

If a second planetary body is perturbed to collide with a pre-existing disk system, then the gaseous debris mass
needed to satisfy the above criteria is about that of Vesta, conservatively speaking.  The largest asteroids in the Solar 
System -- Ceres, Vesta, and Pallas -- are best described as intact planetary embryos, and while their sizes and masses 
are exceptional even among the largest objects in the Main Belt members, such masses may be more commonplace 
among planetesimals formed at the intermediate mass, A- and F-type progenitors of typical, extant white dwarfs.  If the
entire mass is not converted to gas when a second body impacts a disk, and remaining material is gradually delivered 
to the star in a particulate disk, then a Vesta analog represents the minimum mass necessary to account for the data.

In the case where singular events are the cause of metal-enriched white dwarfs, and substantial gas mass is only 
produced immediately following a tidal disruption, the parent bodies must be larger still.  Again for the criteria above, 
one needs 10\% of a $10^{24}$\,g or 1\% of a $10^{25}$\,g parent body to be accreted within a 100\,yr period, as gas, 
with the remaining mass gradually delivered to the star in a particulate disk, at the rates observed for the DAZ stars.
Here the parent bodies would be at least massive as Ceres, and perhaps as large as Pluto.  Also, the particulate disk 
lifetimes necessary to accrete the remaining mass at more modest rates exceed $10^6$\,yr.  Neither of these predictions
is a priori unreasonable, and further observations and modeling may help to constrain these quantities.

\section{Conclusions}

The rapid infall of gaseous debris at white dwarfs can account for the difference between the currently observed and 
ongoing metal accretion rates at DAZ stars and the time-averaged, historical rates inferred for the DBZ stars.  The data
constrain the lifetime of the necessary, high-rate accretion episodes to be strictly less than 10$^4$\,yr, and gas disk 
models with $\alpha\sim0.1$ can achieve rates approaching or exceeding 10$^{15}$\,g\,s$^{-1}$; sufficient to consume 
Vesta in 10\,yr.  The several hundred DA white dwarfs surveyed for metals suggest that the high-rate accretion timescale 
is likely less than $10^3$\,yr, while a search of $\sim10\,000$ DA stars is necessary to confidently detect one in a high 
state that lasts $\sim100$\,yr.  

Such rapid accretion is not unrealistic and within an order of magnitude of the highest accretion rates found for cataclysmic 
variables in quiescence \citep{bas05,gan95}.  Given that {\em ROSAT} was capable of detecting x-ray luminosities as low
as $10^{30}$\,erg\,s$^{-1}$ at distances beyond 100\,pc \citep{pre12}, corresponding to accretion rates of $10^{13}$\,g\,s$^
{-1}$ for typical 0.6\,$M_{\odot}$ white dwarfs, it is likely that near-future facilities will detect the high-rate accretion episodes
predicted here, if the model is accurate (see \citealt{bea12} for alternative predictions).  {\em eROSITA} will be a factor of 100 
times more sensitive than {\em ROSAT} in the $0.5-2$\,keV band\footnote{http://www.mpe.mpg.de/erosita/science.php?lang=en}, 
meaning that accretion rates as low as $10^{11}$\,g\,s$^{-1}$ may be detectable within a few hundred pc.  Moreover, wide-field 
and high-cadence surveys such as LSST and Pan-Starrs may detect optical transients associated with these events.  Lastly, 
{\em GAIA} should be capable of identifying a single white dwarf accreting in a high state, as it should identify $10^6$ white 
dwarfs and thus yield at least 10\,000 metal-polluted stars.  These data will help to constrain $t_h$ and $\dot M_h$.

A readily testable hypothesis is that all relatively warm DAZ stars accreting near or below Poynting-Robertson drag rates 
are accreting from particulate, not gaseous, debris disks.  \citet{far10b} has shown that narrow rings of dust exist at some
polluted white dwarfs, and that 10$^{22}$\,g can easily be missed in photometric surveys for infrared excess.  Those DAZ 
stars without obvious infrared excess in {\em Spitzer} IRAC photometry may reveal dust with higher sensitivity observations.
High signal-to-noise spectroscopy over the $5-15$\,$\mu$m range, where both thermal and silicate emission are salient 
for disks observed to date \citep{rea09,jur09}, is perhaps the ideal choice for the detection of subtle infrared excesses at 
white dwarfs.  This would be somewhat analogous to the identification of dust at HD\,69830 \citep{lis07}, where the 
fractional infrared luminosity is relatively low.

If the accretion of gaseous debris delivers significant mass to metal-contaminated white dwarfs, the destroyed and
devoured parent bodies must be at least as massive as the largest Solar System asteroids, and possibly comparable
in mass to the largest moons.  Such objects will be differentiated, consistent with the few polluted white dwarfs with 
detailed abundance measurements for major elements of planetary solids.  The overall results show that quasi-steady 
state accretion may be important, and models that constrain the frequency of high accretion states will strengthen the
analytical connection between the observed stellar abundances and the chemistry of planetary debris at DBZ and 
very cool DAZ white dwarfs.

\section*{ACKNOWLEDGMENTS}
The authors thank the anonymous referee for feedback which improved the quality and clarity of the manuscript.

\label{lastpage}

\end{document}